%% file: pap.tex
\documentclass[submission,copyright,creativecommons]{eptcs}
\input{macros}
\providecommand{\event}{HCVS 2016}
\begin{document}

\title{Hierarchical State Machines as Modular Horn Clauses
 \thanks{ This work was partially supported by the ANR-INSE-2012
    CAFEIN project and NASA Contract No. NNX14AI09G.
}}

\author{Pierre-Lo\"ic Garoche
\institute{DTIM, UFT, Onera -- The French Aerospace Lab}
\and
Temesghen Kahsai
\institute{NASA Ames / CMU}
\and
Xavier Thirioux
\institute{IRIT/ENSEEIHT, UFT, CNRS}
}
\def\titlerunning{Hierarchical State Machines as Horn Clauses}
\def\authorrunning{P-L. Garoche , T. Kahsai \& X. Thirioux}

\maketitle

\begin{abstract}

\input{abstract}
\end{abstract}

\section{Introduction}
\label{sec:intro}

\input{intro}

\section{Automaton as hierarchical state machines}
\label{sec:automaton}
\input{automaton}

\section{Synchronous dataflow programs as Horn clauses}
\label{sec:compiling_dataflow}
\input{compiling_dataflow}

\section{Compilation of automaton}
\label{sec:compiling_automaton}
\input{compiling_automaton}

\section{Example of encoding}
\label{sec:example}
\input{example}
\section{Conclusion}\label{sec:concl}
\input{conclu}

\bibliographystyle{eptcs}
\bibliography{biblio}



\end{document}

%% file: macros.tex
\usepackage[T1]{fontenc}
\usepackage[utf8]{inputenc}
\usepackage{amsmath,amsfonts,amssymb}
\input{header_tikz}
\usepackage{stmaryrd} 
\usepackage{xspace}
\usepackage{color}
\usepackage{paracol,listings}
\usepackage{newlfont}
\usepackage{wrapfig}
\usepackage[normalem]{ulem} 
\usepackage{amsthm}

\usepackage{multicol}
\usepackage{graphicx}
\usepackage{comment}
\usepackage{url}
\usepackage[noend]{algorithmic}
\usepackage{algorithm}
\usepackage{hyperref}
\usepackage{float}

 \newsavebox{\measurebox}

\newtheorem{defi}{Definition}

\usepackage{lipsum}
\newcounter{examplecounter}

\definecolor{lg}{rgb}{0.9, 0.9, 0.9}

\input{header_lustre}

\newcommand{\lustrec}{\textrm{\textsc{LustreC}}\xspace}

\definecolor{listingComment}{rgb}{0.3,0.7,0.3}
\definecolor{listingString}{rgb}{0.3,0.3,0.7}
\definecolor{listingBackground}{rgb}{0.95,0.95,0.95}



%% file: header_tikz.tex
\usepackage{etex,tikz}
\usetikzlibrary{automata,mindmap,decorations.fractals,shapes,arrows,decorations.pathmorphing,backgrounds,positioning,fit,decorations.pathreplacing,calc,through,shadows,matrix,chains,shadows.blur}



%% file: header_lustre.tex
\usepackage{color,listings}

\lstdefinelanguage{lustre}
    { morekeywords={
        imported, node, 
        bool, int, float,
        initial, let, tel, until, unless, type, var, when, whennot, enum,
        match, if, then, else, state, do, done, resume, restart, returns, merge,
        pre, current, last, map, red, fill, default, every, function,
        fby, automaton, tail, implies, pre, assert, rule, query, var, declare, rel,
        PROPERTY, or, and, requires, ensures, observer, rule, query, Bool, Int, ite, not},
      morecomment=[l][\color{blue}]{--},
      morecomment=[s][\color{blue}]{(*}{*)},
      morecomment=[l][\color{blue}]{assert},
      moredelim=[is][\color{red}]{£}{£}
    }[keywords,comments]

\lstnewenvironment{lustre}
  {\lstset{language=lustre}}{}

\def\lustreinline{\lstinline[language=lustre,basicstyle=\normalsize\sffamily]}

\lstdefinelanguage{horn}
    { morekeywords={
        =>, rule, and, or, declare, datatypes, rel},
      morecomment=[l][\color{blue}]{--},
      morecomment=[s][\color{blue}]{(*}{*)},
      morecomment=[l][\color{blue}]{assert},
      moredelim=[is][\color{red}]{£}{£}
    }[keywords,comments]

\lstnewenvironment{horn}
  {\lstset{language=horn}}{}

\def\horninline{\lstinline[language=horn,basicstyle=\normalsize\sffamily]}

\lstnewenvironment{ccode}
  {\lstset{language=C}}{}

\usepackage{caption}
\usepackage{newfloat}

\DeclareFloatingEnvironment[fileext=lol, name=Listing]{listing}
\usepackage{subcaption}
\DeclareCaptionSubType{listing}


%% file: abstract.tex
In model based development, embedded systems are modeled using a mix of dataflow formalism, that capture the flow of computation, and hierarchical state machines, that capture the modal beahviour of the system. For safety analysis, existing approaches rely on a compilation scheme that transform the original model (dataflow and state machines) into a pure dataflow formalism. Such compilation often result in loss of important structural information that capture the modal behaviour of the system. In previous work we have developed a compilation technique from a dataflow formalism into modular Horn clauses. In this paper, we present a novel technique that faithfully compile hierarchical state machines into modular Horn clauses. Our compilation technique preserves the structural and modal behavior of the system, making the safety analysis of such models more tractable.


%% file: intro.tex
Model-based development is a leading technique in developing software for critical embedded systems such as automotive, avionics systems, train controllers and medical devices. Typically such systems are modeled using a mix of {\it dataflow} formalism and {\it hierarchical state machines}. For instance, Matlab Simulink~\cite{simulink} or Esterel SCADE~\cite{scade} diagrams are typically used to specify aspects of a system that can be modeled by differential equations relating inputs and outputs (i.e., dataflow), while Matlab Stateflow~\cite{stateflow} charts or Esterel SCADE automata usually model the control aspects. The extensive use of the 
aforementioned formalism in the development of safety-critical systems, associated with certification standards~\cite{do178} that recommend the use of formal methods for the specification, design, development and verification of software, makes a formal treatment of these notations extremely crucial. 

For the purpose of safety analysis, Simulink/SCADE models
are compiled to a lower level modeling language, usually a synchronous
dataflow language such as Lustre~\cite{DBLP:conf/popl/CaspiPHP87}. Preserving the original
(hierarchical and modular) structure of the model is paramount to the
success of the analysis process. In~\cite{DBLP:journals/corr/GarocheGK14} we have illustrated a technique to preserve such structure via a modular compilation
process. Specifically, we presented a technique that consists of compiling in a
modular fashion Lustre programs into constrained Horn clauses. In this paper, we extend our previous compilation schema to handle hierarchical state machines (i.e. Stateflow diagrams or SCADE automata). Hierarchical state machines allows to capture the complex {\it modal} behavior of a reactive system. In these systems, the {\it modes} (or {\it state}) of the software drive the behavior of the device. For example, in a car cruise controller, it could be a state machine describing how the controller engages and disengages depending on a number of parameters and actions. 

Existing approaches compile hierarchical state machines into ``pure'' dataflow formalism (such as Lustre). While this approach is rather general, it has the disadvantage that the structure of state machine gets lost in the translation. This can have crucial consequences for verification methods based on inductive arguments, such as {\it k-induction}\cite{DBLP:journals/corr/abs-1111-0372} or {property directed reachability}\cite{pdr}, because the logical encoding ends up creating a state space with states that do not correspond to any state of the original state machine, and so are unreachable by the resulting transition system. These states are problematic because they typically lead to spurious counter-examples for the inductive step of the verification process.

In this paper, we propose a technique to faithfully compile hierarchical state machines expressed as automata in the Lustre language into modular Horn clauses. Our compilation technique preserves the structural and modal behavior of the system, making the analysis of such models more tractable. Specifically, this paper makes the following contributions:

\begin{itemize}
\item a state-preserving encoding of hierarchical state machines as
  pure clocked-dataflow models. This encoding is inspired
  by the work described in ~\cite{DBLP:conf/lctrts/BiernackiCHP08}. Our technique differ in how we encode the state of each automaton, which gives a more flexible encoding.

\item a compilation of hierarchical state machines into modular Horn clauses.
 \item finally, an implementation of the proposed compilation in \lustrec~\cite{lustrec} -- an open source compiler for Lustre programs.
\end{itemize}

The rest of the paper is structured as follows: in the next sub-section, we give an overview of the synchronous dataflow language Lustre. In Section~\ref{sec:automaton} we describe the semantics of the hierarchical state machines that we consider in this paper. In Section~\ref{sec:compiling_dataflow}
we describe our structure preserving compilation scheme. In Section~\ref{sec:compiling_automaton} we illustrate the extensions of the compiler to handle the compilation of automata in to Horn clauses. Finally, in Section~\ref{sec:example} we illustrate
our compilation approach on a simple yet representative example.

\subsection{Background}
Synchronous languages are a class of languages proposed for the design of so
called ``reactive systems'' -- systems that maintain a permanent interaction
with physical environment. Such languages are based on the theory of synchronous
time, in which the system and its environment are considered to both view time
with some ``abstract'' universal clock. In order to simplify reasoning about
such systems, outputs are usually considered to be calculated
instantly~\cite{DBLP:journals/pieee/BenvenisteCEHGS03}. Examples of such languages include
Esterel~\cite{esterel}, Signal~\cite{signal} and
Lustre~\cite{DBLP:conf/popl/CaspiPHP87,lustre2}. In this paper, we will concentrate on the
latter. Lustre combines each data stream with an associated clock as a way to
discretize time. The overall system is considered to have a universal clock
that represents the smallest time span the system is able to distinguish, with
additional, coarser-grained, user-defined clocks. Therefore the overall system
may have different subsections that react to inputs at different frequencies. At
each clock tick, the system is considered to evaluate all streams, so all values
are considered stable for any actual time spent in the instant between ticks. A
stream position can be used to indicate a specific value of a stream in a given
instant, indexed by its clock tick. A stream at position $0$ is in its initial
configuration. Positions prior to this have no defined stream value. A Lustre
program defines a set of equations of the form:
\lustreinline!y$_1$, $\dots$, y$_n$ = f(x$_1$,$\dots$, x$_m$)!
where $y_i$ are output or local variables and $x_i$ are input
variables. Variables in Lustre are used to represent individual streams and they
are typed, with basic types including streams of \textit{Real} numbers,
\textit{Integers}, and \textit{Booleans}. Lustre programs and subprograms are
expressed in terms of \textit{Nodes}. Nodes directly model subsystems in a
modular fashion, with an externally visible set of inputs and outputs. A
\textit{node} can be seen as a mapping of a finite set of input streams (in the
form of a tuple) to a finite set of output streams (also expressed as a
tuple). The \textit{top node} is the main node of the program, the one that
interface with the environment of the program and can never be called by another
node.

At each instant $t$, the node takes in the values of its input streams and
returns the values of its output streams. Operationally, a node has a cyclic
behavior: at each cycle $t$, it takes as input the value of each input stream at
position or instant $t$, and returns the value of each output stream at instant
$t$. This computation is assumed to be immediate in the computation
model. Lustre nodes have a limited form of memory in that, when computing the
output values they can also look at input and output values from previous
instants, up to a finite limit statically determined by the program itself.

Typically, the body of a Lustre node consists of a set of definitions, stream
equations of the form $x = t$ (as seen in Figure~\ref{fig:lustre_synt}) where
$x$ is a variable denoting an output or a locally defined stream and $t$ is an
expression, in a certain stream algebra, whose variables name input, output, or
local streams. More generally, $x$ can be a tuple of stream variables and $t$ an
expression evaluating to a tuple of the same type. Most of Lustre's operators
are point-wise lifting to streams of the usual operators over stream values. For
example, let $x = [x_0, x_1, \dots]$ and $y = [y_0, y_1, \dots]$ be two integer
streams. Then, $x + y$ denotes the stream $[x_0 + y_0; x_1 + y_1, \dots]$; an
integer constant $c$, denotes the constant integer stream $[c,c, \dots]$. Two
important additional operators are a unary shift-right operator \textit{pre}
(``previous''), and a binary initialization operator $\rightarrow$ (``followed
by"). The first is defined as $pre(x) = [u, x_0, x_1, \dots]$ with the value $u$
left unspecified. The second is defined as $x \rightarrow y = [x_0, y_1, y_2,
\dots]$. Syntactical restrictions on the equations in a Lustre program guarantee
that all its streams are well defined: e.g. forbidding recursive definitions
hence avoiding algebraic loops.

\begin{figure}[t]
\centering

\begin{lustre}
type run_mode = enum { Start, Stop };

function switch (mode_in : run_mode) returns (mode_out : run_mode);
let mode_out = if mode_in = Start then Stop else Start; tel

node count (tick:bool) returns (seconds:int);
let seconds = 0 -> pre seconds + 1; tel

node stopwatch (tick:bool; start_stop:bool; reset:bool) returns (seconds : int);
var run : run_mode clock;
let run = Stop -> if start_stop then switch(pre run) else pre run;
    seconds = merge run (Start -> count(tick when Start(run)) every reset)
                        (Stop -> (0 -> pre seconds) when Stop(run));
tel
\end{lustre}


\caption{A simple Lustre program.}
\label{fig:lustre_synt}
\end{figure}

Figure~\ref{fig:lustre_synt} illustrate a simple {\it stopwatch} example using Lustre
{\it enumerated clocks} and {\it node reset}. Enumerated clocks are an advanced form of the traditional Lustre clocks. They allow to sample a value of a flow depending on the value of a
clock. For example, the expression ``\lustreinline!tick when Start(run)! '' denotes
a signal that is only defined when the clock flow run has value Start.  The sampled flows can be gathered together using the {\bf merge} operator as
in the definition of variable ``\lustreinline!seconds!'' in node
\lustreinline!stopwatch!.  
%
%
Moreover, a node call can be reset to its initial state when a given boolean
condition is set to true. For example, in Figure~\ref{fig:lustre_synt} the expression  ``\lustreinline!count(..)! \lustreinline!every reset!'' will return the initial state of the node \lustreinline!count!. The function {\it switch} is a memoryless node, hence is declared with the keyword {\it function}. 



%% file: automaton.tex

Synchronous semantics of hierarchical state machines and their compilation to imperative code has been investigated in different articles, e.g. \cite{Harel:1987:SVF,Harel:1998:MRS,DBLP:conf/concur/UseltonS94,DBLP:conf/fase/HamonR04}, which resulted in a vast number of
different incompatible semantics. Furthermore, the challenges of mixing
state machines and dataflow formalism has also been the subject of intensive
studies, from ~\cite{Maraninchi:1998:Mode}
to~\cite{DBLP:conf/emsoft/ColacoPP05,DBLP:conf/emsoft/ColacoHP06}. In our setting, we follows the
approach developed in \cite{DBLP:conf/emsoft/ColacoPP05,DBLP:conf/emsoft/ColacoHP06}, which is to the best of our
knowledge the most disciplined and simple approach. This technique is also implemented in the commercial KCG
Scade/Lustre compiler~\cite{scade}.


\begin{figure}
\centering
\resizebox{.8\linewidth}{!}{\input{automaton_logics.tikz}}
\caption{Automaton as a pure dataflow}
\label{automaton}
\end{figure}
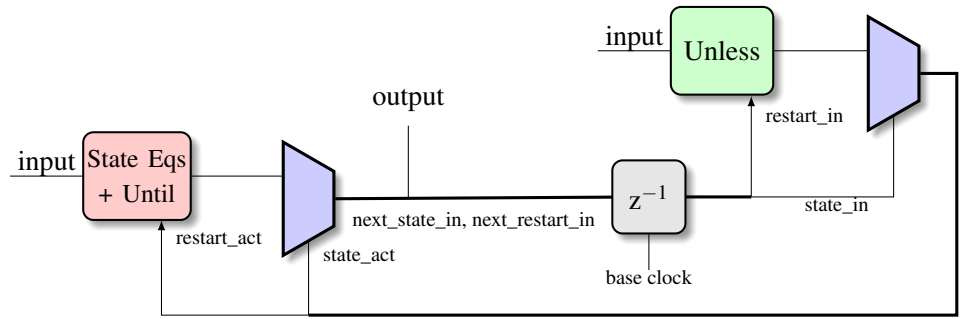

Informally, the modular compilation scheme developed in
~\cite{DBLP:conf/emsoft/ColacoPP05,DBLP:conf/emsoft/ColacoHP06} is enforced at
the expense of raw (and somehow undesired) expressivity, disallowing for
instance transitions that go through boundaries of hierarchical state machines
or the firing of an unbounded number of transitions per instant (e.g. in Matlab
Simulink and Stateflow). For instance, in Figure~\ref{automaton}, at each
instant, two pairs of variables are computed: a putative $state\_in$ and an
actual state $state\_act$ and also, for both states, two booleans $restart\_in$
and $restart\_act$, that tell whether their respective state equations should be
reset before execution. The actual state is obtained via a strong ({\tt unless})
transition from the putative state, whereas the next putative state is obtained
via a weak ({\tt until}) transition from the actual state. Only the actual state
equations are executed at each instant. Finally, a reset function is driven by
the {\tt restart}/{\tt resume} keyword switches. As transition-firing conditions
may have their own memories, they can be reset if needed before being
evaluated. Specifically, \texttt{unless} conditions are reset according to
$restart\_in$, whereas \texttt{until} and state equations altogether are reset
according to $restart\_act$.
To recapitulate, a transition is evaluated as follows: unless conditions of the
initial active state -- the putative state -- are evaluated. In case of a valid
one, we jump to the associated state. Then, the state equations are evaluated:
either the ones of the putative state in case of no unless transitions
activated, or the ones of the new state obtained. Then, as a last step, the
until transition of the active state are evaluated and characterize the next
state for the following transition. At most one unless and one until transitions
are evaluated, in this order, at each time step.

Our approach, builds on top of the aforementioned compilation scheme. In our
setting, we promote the computation of strong transition, state equations and
weak transition to independent auxiliary Lustre nodes. This allows a certain
flexibility: (i) independent scheduling and optimization of different state
equations; (ii) addition of code contracts to different states. Those features
are not supported by the commercial KCG suite. Another benefit of our approach
is that we don't modify state equations to take clock constraints, nor local
variables or the reset operation.
Local state information is only recovered
through clock calculus and is not structural any more, as generated code may be
optimized and scattered. We rather only encapsulate state equations in new node
definitions and generate new equations for calling these nodes, greatly
facilitating the management of local state invariants for instance.  Yet, this
comes at the expense of a rather limited loss in expressivity, due to possible
causality issues\footnotemark. Note that inlining these auxiliary nodes is
already an available option that fully recovers the original semantics.

\footnotetext{We recall that the classical causality analysis in modern Lustre
  doesn't cross boundaries of nodes, hence the conservative rejection of some
  correct programs.}

We illustrate in Listings.~\ref{sched:failure}, \ref{sched:solution} and
\ref{sched:causality}, the differences between our approach
and~\cite{DBLP:conf/emsoft/ColacoPP05}, from a user's
viewpoint. Example~\ref{sched:failure} is a typical program that cannot be
statically scheduled and produces a compilation error in both approaches. A
solution may be devised as in Example~\ref{sched:solution}, using an automaton
to encode the boolean switch $i=0$. Even if scheduling is done prior to other
static analyses (and thus is unaware of exclusive automaton states for
instance), we succeed in generating a correct code whereas KCG
fails. Example~\ref{sched:causality} is non-causal and won't compile if we
remove the \textcolor{red}{pre} occurring in the \texttt{unless} clause. But if
we keep it, KCG will handle it correctly whereas our causality analysis will
reject this program. Generally speaking, we forbid \texttt{unless} clauses that
would refer to putative state memories (such as $o$). Accepting these clauses
appear problematic or at least confusing as it makes the putative state visible
and distinct from the actual state, thus duplicating state variables.

\begin{listing}
\begin{sublisting}[b]{.5\linewidth}
\begin{lustre}
node failure (i:int) returns (o1, o2:int);
let
  (o1, o2) = if i = 0
             then (o2, i)
             else (i, o1);
tel
\end{lustre}
\vspace{-1em}
\caption{\label{sched:failure}Scheduling failure}
\end{sublisting}
\begin{sublisting}{.5\linewidth}
\vspace{3em}
\begin{lustre}
node solution (i:int) returns (o1, o2:int);
let
  automaton condition
  unless i <> 0 resume KO
  state OK:
  let
    (o1, o2) = (o2, i);
  tel
  state KO:
  unless i = 0 resume OK
  let
    (o1, o2) = (i, o1);
  tel
tel
\end{lustre}
\vspace{-1em}
\caption{\label{sched:solution}Automaton based solution}
\end{sublisting}
%
\\[-2.9cm]
\begin{sublisting}{.5\linewidth}
\begin{lustre}
node triangle (r:bool) returns (o:int);
let
  automaton trivial
  state One:
  unless r || £pre£ o = 100
  let
    o = 0 -> 1 + pre o;
  tel
tel
\end{lustre}
\vspace{-1em}
\caption{Causality issues}
\label{sched:causality}
\end{sublisting}
\caption{Examples comparing our approach with the one developed in~\cite{DBLP:conf/emsoft/ColacoPP05}}

\end{listing}

As a result our encoding is not strictly comparable with Scade or Lustre v6
automaton. For instance, we are unable to type check and compile automaton with memories
within unless conditions. This is possible in our setting. 
%
In summary, our encoding does not flatten the automaton into a single Lustre
node but preserves the structure by associating a Lustre node for each automaton
state. This structure preserving encoding enables us to analyze these models and
compute local invariants associated to automaton states.


%% file: automaton_logics.tikz

{\rm
\begin{tikzpicture}[
 outer sep=0pt,
 inner sep=0pt, 
every node/.style={transform shape,rounded corners,thick},
]

\draw[very thick,sharp corners,fill=blue!20,blur shadow={shadow blur steps=5,shadow blur extra rounding=1.3pt}] (0,0)coordinate (O1) --++(30:.8) coordinate (A1)  --++(90:.75cm)coordinate (B1)  --++(150:.8)coordinate (C1) --cycle;
\node (output1) at ($(A1)!0.5!(B1)$) {};
\node (cmd1) at ([shift={(0,0cm)}]$(O1)!0.5!(A1)$)  {} ;
\node (input1) at ($(C1)! 0.3 !(O1)$) {};

\node [text centered,draw,minimum height=1.2cm,minimum width=1.4cm,text width=1.2cm,fill=green!20,blur shadow={shadow blur steps=5,shadow blur extra rounding=1.3pt}] (unless) at ([shift={(-2cm,0)}]input1) {Unless};

\node [draw,minimum height=1cm,minimum width=1cm,text width=1cm,text centered,fill=gray!20,blur shadow={shadow blur steps=5,shadow blur extra rounding=1.3pt}] (zmun) at ([shift={(-1cm,-2cm)}]unless) {z$^{-1}$};

\node (mult2) at ([shift={(-5cm,0)}]zmun) {};

\node (D2) at ([shift={(0cm,0)}]mult2) {};
\draw[very thick,sharp corners,fill=blue!20,blur shadow={shadow blur steps=5,shadow blur extra rounding=1.3pt}] (D2) --++(-90:.8) coordinate (O2) node {}--++(30:.8) coordinate (A2) node {} --++(90:.75cm)coordinate (B2)  --++(150:.8) coordinate (C2)--cycle;
\node (output2) at ($(A2)!0.5!(B2)$) {};
\node (cmd2) at ([shift={(0,0cm)}]$(O2)!0.5!(A2)$)  {} ;
\node (input2) at ($(C2)! 0.3 !(O2)$) {};

\node [text centered,draw,minimum height=1.2cm,inner sep=1pt,minimum width=1.4cm,text width=1.4cm,fill=red!20,blur shadow={shadow blur steps=5,shadow blur extra rounding=1.3pt}] (until) at ([shift={(-2cm,0)}]input2) {\small State Eqs + Until};

\draw ([shift={(-1cm,0)}]unless.west) --node[above] {input} (unless.west);
\draw (unless.east) -- (input1.west);

\draw ([shift={(-1cm,0)}]until.west) --node[above] {input} (until.west);
\draw (until.east) -- (input2.west);

\draw[very thick] (output2.east) -- node [below=.2cm]{\scriptsize next\_state\_in, next\_restart\_in}(zmun);

\draw ([shift={(0cm,-.5cm)}]zmun.south) node[below] {\scriptsize base clock} -- (zmun);

\draw[very thick] (output1.east) |-  ([shift={(.5cm,0)}]output1.east) |-  ([shift={(0cm,-1cm)}]cmd2.south) ;

\draw[-latex] ([shift={(0cm,-1cm)}]cmd2.south) -|  ([shift={(-.4cm,0)}]until.south east) node[right=.2cm, pos=.9] {\scriptsize restart\_act};
\draw ([shift={(0cm,-1cm)}]cmd2.south) --  (cmd2.south) node[right=.2cm, pos=.8] {\scriptsize state\_act};

\node (anchorunless) at ([shift={(-.3cm,0cm)}]unless.south east) {};
\draw[very thick] (zmun) -- (zmun -| anchorunless);  
\draw (zmun -| anchorunless) -| (cmd1.south) node [pos=.3, below] {\scriptsize state\_in};
\draw[-latex] (zmun -| anchorunless) -| (anchorunless) node [pos=.9, right=.2cm] {\scriptsize restart\_in};

\draw ([shift={(1cm,0)}]output2.east) -- ([shift={(1cm,1cm)}]output2.east) node[above=.2cm] {output};

\end{tikzpicture}
}

%% file: compiling_dataflow.tex
\begin{figure}
\centering
\includegraphics[scale=0.4]{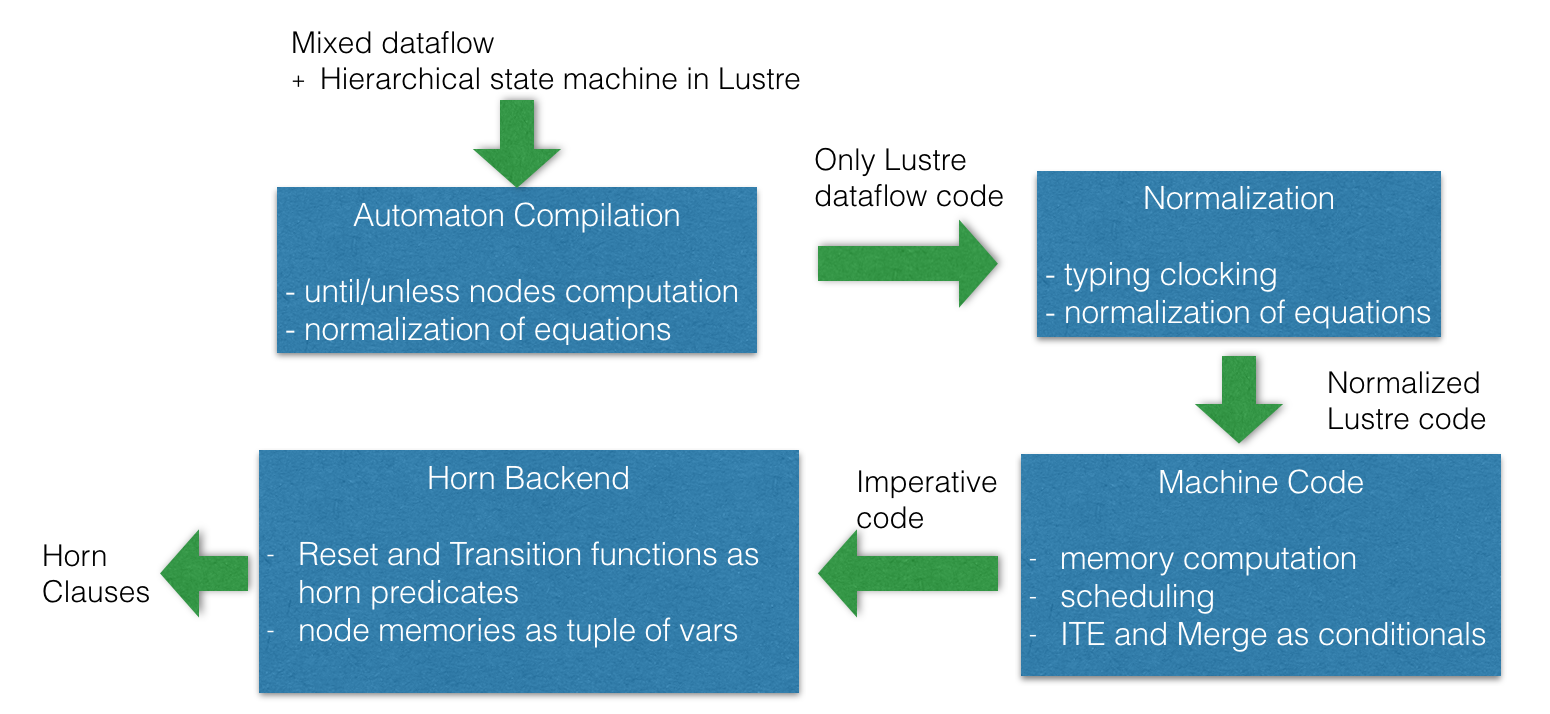}
\caption{Compilation Stages}
\label{fig:compilation_stages}
\end{figure}
In~\cite{DBLP:journals/corr/GarocheGK14} we have developed a compilation technique that translate Lustre programs into modular Horn clauses. In order to accommodate the compilation of automaton we have updated such compilation scheme. In this section, we briefly describe the different stages of the compilation process. For a formal treatment of the compilation stages the reader can refer to ~\cite{DBLP:journals/corr/GarocheGK14}. Figure~\ref{fig:compilation_stages} illustrate the compilation stages implemented in \lustrec.

\paragraph{Automaton compilation.} The first phase, which is new w.r.t. \cite{DBLP:journals/corr/GarocheGK14}, compiles the hierarchical
states machines as pure dataflow expression in Lustre. A detailed description of this phase will be presented in Section.~\ref{sec:compiling_automaton}. 

\paragraph{Normalization.}
This phase infers types and clock for each signal. Each expression is
recursively normalized: node calls, \lustreinline!pre! constructs, tuples, \ldots are
defined as fresh variables. Each function call $foo(args)$ is associated to a unique
identifier $foo^{uid}(args)$. Once normalized, no function calls occurs within
expressions, nor arrows nor definitions of memories through a \texttt{pre}
construct.

\paragraph{Machine code.} In the main compiler process, intended to generate
embedded code, the next phase generates the machine (or imperative) code: 
its flows definitions are replaced
by an ordered sequence of imperative statements. The state of a node instance is
characterized by its memories, i.e. expressions defining memories such as
\lustreinline!x = pre e!, and the node instances that appear in its expressions such as
\lustreinline!x = foo($\ldots$)!. 

\paragraph{Generation of machine code.} At this stage a machine code is generated. 
This amount to replace the flow definitions with an ordered sequence of imperative statements. The state of a node instance is
characterized by its memories, i.e. expressions defining memories such as
$x\ =\ pre\ e$, and the node instances that appear in its expressions such as
$x\ =\ foo(\ldots)$. This tree-like characterization of a node state enable a modular definition of a state as a tree of local memories. 

\begin{defi}[Node memories and instances]
Let $f$ be a Lustre node with normalized equations $eqs$. Then we define its set
of memories and node callee instances as:
\[\begin{array}{rl}
  Mems(f) =& \{ x \mid x = pre\  \_ \in eqs \}\\
  Insts (f) =& \{ (foo, uid) \mid \_ = foo^{uid} (\_) \in eqs \} 
\end{array}
\]
\end{defi}

\begin{figure}
  \begin{subfigure}{.33\textwidth}
    \centering
    \resizebox{!}{3cm}{\input{memory_reset.tikz}}
    \caption{Tree type: memories and instances}
  \end{subfigure}
%
  \begin{subfigure}{.33\textwidth}
\centering
    \resizebox{!}{3cm}{\input{memory_reset2.tikz}}
    \caption{Example of state for node foo}
    \label{fig:mem2}
  \end{subfigure}
\hfill
  \begin{subfigure}{.33\textwidth}
\centering
    \resizebox{!}{3cm}{\input{memory_reset3.tikz}}
\caption{State (\ref{fig:mem2}) after reset}
\label{fig:mem3}
  \end{subfigure}
\caption{Memory trees and reset}
\label{fig:memories}
\end{figure}
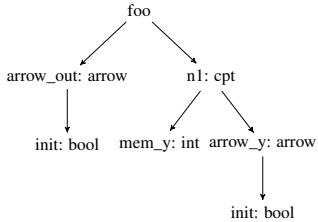
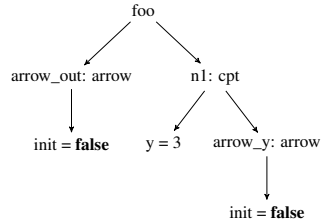
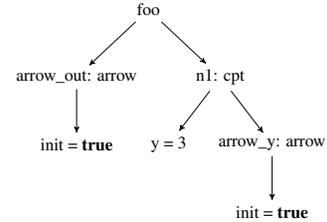

The \textit{follow by} ($\rightarrow$) operator is interpreted as a node instance of a generic
polymorphic node \lustreinline!arrow! as illustrated in Listing~\ref{fig:arrow}.
\begin{listing}
  \begin{sublisting}{.5\linewidth}
\begin{lustre}
node arrow (e1, e2: 'a) returns (out: 'a)
var init: bool;
let
  init = true -> false;
  out = if init then e1 else e2;
tel
\end{lustre}
\vspace{-1em}
\caption{Polymorphic arrow node}
\label{fig:arrow}
\end{sublisting}
\begin{sublisting}{.5\linewidth}
\begin{lustre}
node cpt (z: bool) returns (y: int);
let y = 0 -> if z then 0 else pre y + 1; tel

node foo (z: bool) returns (out: int)
let out = 1 -> foo(z); tel
\end{lustre}
\caption{Simple example with two nodes}  
\label{lst:memory_ex}
\end{sublisting}
\caption{Memories in Lustre}
\end{listing}
Therefore the initial state for a node associated to all its arrow instances and
all its child arrow instances the value true to the memory
\lustreinline!init!. Similarly the activation of a node reset using the operator
\textit{every} modifies the state of this instance of the node \lustreinline!foo! by resetting its arrow
init variable to their initial value \lustreinline!true!, e.g. 
\lustreinline!e = foo (...) every ClockValue(clock_var).!
Figure~\ref{fig:memories} shows the computed memories of the node in Listing~\ref{lst:memory_ex}.

\paragraph{Horn backend.}
At this stage the Horn clauses are generated. The hierarchy of memories and node instances are flattened
and modeled as a tuple of memories. We denote as
\horninline!state$^{\texttt{label}}$ (f, uid)! the tuple of variables denoting
the state of the instance \horninline!uid! of a node \horninline!f!. Different labels are
used to differentiate between different versions of the same variable. We use the labels $c$ and $n$ to denote the (c)urrent and (n)ext value of a memory
\horninline!x!: \horninline!x$^\texttt{c}$! and \horninline!x$^\texttt{n}$!. The internal node {\it arrow} is fitted with a specific {\it reset} rule: 

\begin{horn}[mathescape=true]
              rule (=> (= init$^\texttt{n}$ true) (arrow_reset (init$^\texttt{c}$, init$^\texttt{n}$)))
\end{horn}

\noindent Note that its
state is only defined by the \horninline!init! variable.  
In the proposed encoding, the predicates defining the program semantics 
should enable the reset of a node state as performed in
Fig.~\ref{fig:mem3}. This leads to the following encoding for the reset
function:
\begin{horn}[mathescape=true] 
rule (=> ($\displaystyle\bigwedge_{\texttt{mem}\ \in\ \texttt{Mems(f)}}$ (= mem$^\texttt{n}$  mem$^\texttt{c}$) $\displaystyle\bigwedge_{\texttt{(g, uid)}\ \in\ \texttt{Inst(f)}}$ g_reset (state$^\texttt{c}$ (g,guid), state$^\texttt{n}$ (g,guid)))
         (f_reset (state$^\texttt{c}$(f,uid), state$^\texttt{n}$(f,uid))))
\end{horn}

\noindent The collecting semantics definition of~\cite{DBLP:journals/corr/GarocheGK14} is
modified to rely on \lustreinline!f_reset! instead of \lustreinline!f_init!. It builds the set
of reachable states ({\it Reach}):

\begin{minipage}{\linewidth}
\begin{horn}
rule (=> (f_reset (state$^\texttt{c}$(f,uid), state$^\texttt{n}$(f,uid))) (Reach (state$^\texttt{n}$(f,uid))))
rule (=> (and (f_step (inputs, outputs, state$^\texttt{c}$(f,uid), state$^\texttt{n}$(f,uid)))
              (Reach (state$^\texttt{c}$(f,uid))))
         (Reach (state$^\texttt{n}$(f,uid))))
\end{horn}
\end{minipage}



%% file: memory_reset.tikz
\begin{tikzpicture}[->,>=stealth',
	level 1/.style={sibling distance = 3.2cm, level distance = 1.5cm},
	level 2/.style={sibling distance = 2.3cm, level distance = 1.5cm}, 
	] 
\node (root) {foo}
	child {node {arrow\_out: arrow}
	child {node {init: bool}}}
	child {node {n1: cpt}
    	child {node {mem\_y: int}}
  	    child {node {arrow\_y: arrow}
			child {node {init: bool}}}};
\end{tikzpicture}

%% file: memory_reset2.tikz
\begin{tikzpicture}[->,>=stealth',
	level 1/.style={sibling distance = 3.2cm, level distance = 1.5cm},
	level 2/.style={sibling distance = 2.3cm, level distance = 1.5cm}, 
] 
\node (root2) {foo} 
	child {
          node {arrow\_out: arrow}
	  child {node {init = \textbf{false}}}
	}	
	child {
          node {n1: cpt}
       	  child {node {y = 3}}
  	  child {node {arrow\_y: arrow}
	         child {node {init = \textbf{false}}}	
}};
\end{tikzpicture}

%% file: memory_reset3.tikz
\begin{tikzpicture}[->,>=stealth',
	level 1/.style={sibling distance = 3.2cm, level distance = 1.5cm},
	level 2/.style={sibling distance = 2.3cm, level distance = 1.5cm}, 
] 
\node (root3) {foo} 
	child {
          node {arrow\_out: arrow}
	  child {node {init = \textbf{true}}}
	}	
	child {
          node {n1: cpt}
       	  child {node {y = 3}}
  	  child {node {arrow\_y: arrow}
	         child {node {init = \textbf{true}}}	
}};
\end{tikzpicture}

%% file: compiling_automaton.tex

In this section, we describe the compilation scheme from automaton to modular Horn clauses. This is performed in two stages: (i) compilation of automaton into clocked expressions and (ii) compilation of clocked expressions into Horn clauses.

\subsection{From automaton to clocked expressions} \label{sec:auto_as_clocls}
We denote with $ReadEqs_i$ and $WriteEqs_i$ the set of read and write variables occurring in equations of an automaton state $S_i$. We also denote as $ReadUnless_i$ and $ReadUntil_i$ the set of variables in \texttt{unless} and \texttt{until} clauses.

 Our compilation scheme from automaton to clocked expressions follows Figure~\ref{automaton} and is applied to a generic automaton such as the one described in Figure~\ref{aut:skeleton} (node {\it nd}). As illustrated in Listing~\ref{aut:transitions}, the variables {\it state\_act} and {\it state\_in} are modelled as clocks of enumerated type. 
 Also, two new nodes are introduced for each of the automaton state: one to express the semantics of state equations; and another one to capture the  {\it weak} and {\it strong} transitions (as explained in Section~\ref{sec:automaton}).

Figure~\ref{aut:calling} illustrate the compiled node $c\_nd$ that replace the original
automaton description of node $nd$. Evaluation of each single node call embedding state equations and transitions only takes place 
when its corresponding clock is active; this is done via ``\textbf{when} $Value$($clock$)'' sampling operators applied to all node arguments.

All the node calls that corresponds to the global evaluation of the automaton are then gathered in two \textbf{merge} constructs, which are driven by
the putative state clock {\it state\_in} (for strong transitions) and the actual state
clock {\it state\_act} (for weak transitions and state equations).

\begin{listing}
\begin{sublisting}{.5\linewidth}
\begin{lustre}
node $nd$ ($inputs$) returns ($outputs$);
var $locals$
let
  $other\_equations$
  automaton $aut$
  ...
  state $S_i$:
  ...
  unless $(sc_j, sr_j, SS_j)$
  ...
  var $locals_i$
  let
    $equations_i$
  tel
  ...
  until $(wc_j, wr_j, WS_j)$
  ...
tel
\end{lustre}
\vspace{-1em}
\caption{\label{aut:skeleton}Automaton skeleton}
\end{sublisting}
\begin{sublisting}{.5\linewidth}
\begin{lustre}
type $aut$_type = enum { $S_1$, ..., $S_n$ };
...
node $S_i$_unless ($ReadUnless_i$)
returns (restart_act : bool, 
         state_act : $aut$_type clock);
let
  (restart_act, state_act) =
    if $sc_1$ then ($sr_1$, $SS_1$) else
    if $sc_2$ then ($sr_2$, $SS_2$) else
    ...
    (false, $S_i$);
tel

node $S_i$_handler_until ($ReadEqs_i \cup ReadUntil_i$)
returns (restart_in : bool, 
         state_in : $aut$_type clock,
         $WriteEqs$);
var $locals_i$
let
  (restart_in, state_in) =
    if $wc_1$ then ($wr_1$, $WS_1$) else
    if $wc_2$ then ($wr_2$, $WS_2$) else
    ...
    (false, $S_i$);
  $equations_i$
tel
\end{lustre}
\vspace{-1em}
\caption{\label{aut:transitions}Clocked expression as new nodes.}
\end{sublisting}
\caption{Automaton in Lustre and their representation as clocked expressions in Lustre nodes.}
\end{listing}

\begin{listing}[t!]
\begin{lustre}
node $c\_nd$ ($inputs$) returns ($outputs$);
var $locals$; 
    $aut$_restart_in, $aut$_next_restart_in, $aut$_restart_act : bool;
    $aut$_state_in, $aut$_next_state_in, $aut$_state_act : $aut$_type clock;
let
  $\dots$
  ($aut$_restart_in, $aut$_state_in) =
    (false, $S_1$) -> pre ($aut$_next_restart_in, $aut$_next_state_in);
  ($aut$_restart_act, $aut$_state_act) = merge $aut$_state_in
     $\dots$
     ($S_i$ -> $S_i$_unless(($ReadUnless_i$) when $S_i$($aut$_state_in)) every $aut$_restart_in)
     $\dots$
  ($aut$_next_restart_in, $aut$_state_next_in, $WriteEqs$) = merge $aut$_state_act
     $\dots$
     ($S_i$ -> $S_i$_handler_until(($ReadEqs_i \cup ReadUntil_i$) when $S_i$($aut$_state_act)) every $aut$_restart_act)
    $\dots$
tel
\end{lustre}
\vspace{-2em}
\caption{\label{aut:calling}Compiled node $c\_nd$ from node $nd$ in Figure~\ref{aut:skeleton}.}
\end{listing}

\subsection{Compiling clocked expressions into modular Horn clauses}
\label{sec:clocks_in_horn}

Once the automaton structure as been compiled into clocked expression, the second step is to encode them as Horn clauses. Here, we use the Horn clause format introduced in Z3~\cite{pdr}, where
$(\mathtt{rule}\ expr)$ universally quantify the free variables of the SMT-LIB
expression $expr$. At this level, the challenges are to be able to express within the Horn formalism the following concepts: (i) the clock's feature of
Lustre, (ii) the reset functionality of a node, (iii) the declaration of enumerated clocks, (iv) clocked expression with the
\textbf{when} operator, (v) merge of clocked expressions and reset of node state on
conditionals with the \textbf{every} operator. In the following we illustrate how we capture using Horn clauses the above mentioned concepts. 
\newpage
Clock values are defined as regular enumerated type in SMT-LIB format:

\begin{horn}
                 (declare-datatypes () ((clock_type Start Stop)))
\end{horn}

\noindent Combination of {\bf merge} and {\bf when} operator are required for the clock calculus
(i.e. clock typing) but are ignored when generating the final code. Merge
constructs act as a switch-case statement over well-clocked expression. For example, the following well-clocked Lustre expression:

\begin{lustre}
      seconds = merge run (Start -> x when Start(run)) (Stop -> y when Stop(run))
\end{lustre}

is translated to the imperative switch-case expression:

\begin{ccode}
           switch (run) { case Start : e = x; break;
                          case Stop  : e = y; break }
\end{ccode}

\noindent Since each case definition is purely functional, this can be directly expressed as the following constraint:

\begin{horn}
                (and (=> (= run Start) (= e x)) 
                     (=> (= run Stop)  (= e y)))
\end{horn}
%
%
The next item is to capture the reset of a node's state using the {\bf every} operator,\\ e.g. \lustreinline!count (x) every condition!. During the compilation process, such expression generate a machine code instruction:

\begin{ccode}
                     if (condition) { Reset(count, uid) };
\end{ccode}

which gets translated to an imperative statement:

\begin{ccode}
                     if (condition) { count_reset(state_count_uid) } else {};
\end{ccode}

where {\it state\_count\_uid} is a {\it struct} that denotes a node's state instance. s
This conditional statement will perform a side-effect update of the memory state
and impact the computation of the next state and outputs. How do we capture this in the Horn encoding? Let us first look how we encode a step transition in Horn clauses. A step transition is basically a
relationship (i.e. a predicate) between inputs, outputs, previous state and next state.

\begin{horn}
    relationship(inputs, outputs, old_state, new_state) = (and (...))
\end{horn}

\noindent Typically this relationship would be used to define the step transition as follows:

\begin{horn}[mathescape=true] 
rule (=> (relationship(inputs, outputs, state$^\texttt{c}$(f,uid), state$^\texttt{n}$(f,uid)))
         (f_step (inputs, outputs, state$^\texttt{c}$(f,uid), state$^\texttt{n}$(f,uid))))
\end{horn}

For resetting the node's state, the value used for the state is 
\horninline!f_reset(state$^\texttt{c}$(f,uid))! (instead of
\horninline!state$^\texttt{c}$(f,uid)!). In addition to the two state labels
used to denote current $c$ and next value $x$, we introduce an intermediate
label $i$. In case of a transition without reset, the intermediate version would
be directly defined as the current one.

\begin{horn}[mathescape=true] 
rule (=> (and (= (state$^\mathtt{i}$(f,uid)) (state$^\mathtt{c}$(f,uid)))
              (relationship(inputs, outputs, state$^\texttt{i}$(f,uid), state$^\texttt{n}$(f,uid))))
         (f_step (inputs, outputs, state$^\texttt{c}$(f,uid), state$^\texttt{n}$(f,uid))))
\end{horn}

\noindent Finally, the reset function is encoded as follows:
%

\begin{horn}[mathescape=true] 
rule (=> (and (= (state$^\mathtt{i}$(f,uid)) 
                 (if condition then f_reset(state$^\mathtt{c}$(f,uid)) else state$^\mathtt{c}$(f,uid)))
              (relationship(inputs, outputs, state$^\texttt{i}$(f,uid), state$^\texttt{n}$(f,uid))))
         (f_step (inputs, outputs, state$^\texttt{c}$(f,uid), state$^\texttt{n}$(f,uid))))
\end{horn}

%% file: example.tex
As an example of the proposed compilation process, we consider a simple
Lustre program that compares three implementations of a 2-bit counter: a low-level
Boolean implementation, a higher-level implementation using integers and an automaton based counter. The
\texttt{greycounter} node  (cf. Fig.~\ref{fig:greycounter}) internally repeats the sequence $ab = \{00, 01, 11,
10,00, . . .\}$ indefinitely, while the \texttt{integercounter} node (cf. Fig.~\ref{fig:intloopcounter}) repeats the
sequence $time = \{0, 1, 2, 3, 0, \dots \}$. The automaton based node (cf. Fig.~\ref{fig:automaton_lustre}) is a state machines with 4 states and it basically alternates between them.
\begin{listing}
  \begin{minipage}[b]{.33\textwidth}
\begin{sublisting}{\linewidth}
\begin{lustre}
node auto (x:bool) returns (out:bool);
let
  automaton four_states
  state One : 
  let 
   out = false;
  tel until true restart Two
  state Two : 
  let 
   out = false;
  tel until true restart Three
  state Three : 
  let 
   out = true;
  tel until true restart Four
  state Four :
  let 
   out = false;
  tel until true restart One
tel
\end{lustre}
\caption{Automaton-based counter}
\label{fig:automaton_lustre}
\end{sublisting}
  \end{minipage}
\hspace{2cm}
\begin{minipage}[b][4cm]
[s]
{.33\textwidth}
\centering
\begin{sublisting}{\linewidth}
\begin{lustre}
node greycounter (x:bool) returns (out:bool);
var a,b:bool;
let
  a = false -> not pre(b);
  b = false -> pre(a);
  out = a and b;
tel
\end{lustre}
\vspace{-1em}
\caption{Boolean-based counter}
\label{fig:greycounter}
\end{sublisting}

\vspace{1cm}

\begin{sublisting}{\linewidth}
\begin{lustre}
node intloopcounter (x:bool) returns (out:bool);
var time: int;
let
  time = 0 -> if pre(time) = 3 then 0
              else pre time + 1;
  out = (time = 2);
tel
\end{lustre}
\vspace{-1em}
\caption{Integer-based counter}
\label{fig:intloopcounter}
\end{sublisting}
\end{minipage}
\caption{Automaton-(\ref{fig:automaton_lustre}), Boolean-(\ref{fig:greycounter}) and Integer-(\ref{fig:intloopcounter}) based implementation of a 2-bit counter.}
\end{listing}

\noindent In the first phase of our compilation scheme a clock is generated to encode the automaton states:

\begin{lustre}
                    type auto_ck = enum {One, Two, Three, Four };  
\end{lustre}
 

\noindent Each automaton state is associated to a stateless function describing respectively its strong ({\it unless})
transitions and its weak ({\it until}) ones. 
\newpage
\noindent To keep the presentation simpler we present the encoding of the state {\it Four}:
%
%
%
\begin{lustre}
     function Four_handler_until (restart_act: bool; state_act: auto_ck) 
        returns (restart_in: bool; state_in: auto_ck; out: bool)
      let -- encodes the next state, here One  
          restart_in, state_in = (true,One);  
          out = false;-- returns true in the handler for state Three
      tel
\end{lustre}

\noindent The handler and the {\it until} function assigns the next state to the state {\it One} and require the node to be restarted. Listing~\ref{ls:automatonless} shows the generated Lustre node without an automaton. 
\begin{listing}
\begin{sublisting}{\linewidth}
\begin{lustre}
   node auto (x: bool) returns (out: bool)
     var mem_restart: bool; mem_state: auto_ck;  
       four_restart_in: bool; four_state_in: auto_ck; four_out: bool ;
       four_restart_act: bool; four_restart_in: auto_ck;
       ... -- similar declarations for other states
       next_restart_in: bool; restart_in: bool;
       restart_act: bool; next_state_in: auto_ck;
       state_in: auto_ck clock; state_act: auto_ck clock;
    let   
      restart_in, state_in = ((false,One) -> (mem_restart,mem_state));
      mem_restart, mem_state = pre (next_restart_in,next_state_in);
      next_restart_in, next_state_in, out = 
               merge state_act (One -> ...) (Two -> ...)  (Three -> ...) 
                               (Four -> (four_restart_in, four_state_in, four_out));
      four_restart_in, four_state_in, four_out = 
      Four_handler_until (     restart_act when Four(state_act),
                               state_act when Four(state_act)) 
                          every (restart_act);
       ... -- similar definitions for other states
      restart_act, state_act = merge state_in (One -> ...) 
                                              (Two -> ...) 
                                              (Three -> ...) 
                                        (Four -> (four_restart_act, four_state_act));
      four_restart_act, four_state_act = 
        Four_unless (     restart_in when Four(state_in),
                          state_in when Four(state_in)) 
                     every (restart_in);
      ... -- similar definitions for other states 
     tel
\end{lustre}
\vspace{-1em}
\end{sublisting}
\caption{\label{ls:automatonless}Generated Lustre code without automaton.}
\end{listing}

\noindent The next stage of the compiler produces the Horn clauses. Enumerated type enable the declaration of clock's values:

\begin{horn}
                   (declare-data auto_ck () ((auto_ck One Two Three Four)))
\end{horn}

\noindent The functions for {\it until} and {\it unless} are defined as Horn predicates ({\it Four\_handler\_until} and {\it Four\_unless} respectively) in the following way:

\begin{horn}
       (declare-rel Four_handler_until (Bool auto_ck Bool auto_ck Bool))
       (rule (=> (and (= out false) (= state_in One) (= restart_in true))
                 (Four_handler_until restart_act state_act restart_in  state_in out)))
       (declare-rel Four_unless (Bool auto_ck Bool auto_ck))
       (rule (=> (and (= state_act state_in) (= restart_act restart_in))
                 (Four_unless restart_in state_in restart_act state_act)))
\end{horn}
\newpage
Finally the reset ({\it auto\_reset}) and step ({\it auto\_step}) predicates are defined as follows respectively:

\begin{horn}
(rule (=> (and (= mem_restart_m mem_restart_c) (= mem_state_m mem_state_c) 
               (= arrow.init_m true))
      (auto_reset mem_restart_c mem_state_c arrow.init_c
                  mem_restart_m mem_state_m arrow.init_m)))

(rule (=> 
    (and (= arrow.init_m arrow.init_c)
         (= arrow.init_x false) -- update of arrow state
    (and (=> (= arrow.init_m true) -- current arrow is first 
             (and (= state_in One)
                  (= restart_in false)))
         (=> (= arrow.init_m false) -- current arrow is not first
             (and (= state_in mem_state_c)
               (= restart_in mem_restart_c))))
     (and (=> (= state_in Four) -- unless block for automaton state Four
          (and (Four_unless restart_in state_in four_restart_act four_state_act)
               (= state_act four_state_act)
               (= restart_act four_restart_act)))
             ...) -- similar definition for other states       
     (and (=> (= state_act Four) -- handler and until block for state Four
          (and (Four_handler_until restart_act state_act 
                                   four_restart_in four_state_in four_out)
              (= out four_out)
              (= next_state_in four_state_in)
              (= next_restart_in four_restart_in)))
               ...) -- similar definition for other states       
          (= mem_state_x next_state_in) -- next value for memory mem_state
          (= mem_restart_x next_restart_in)) -- next value for memory mem_restart
(auto_step x    -- inputs
           out  -- outputs
           mem_restart_c mem_state_c arrow.init_c    -- old state
           mem_restart_x mem_state_x arrow.init_x))) -- new state
\end{horn}

\noindent Once the Horn clauses are generated, a Horn clause solver can be used to perform verification and/or testing. For example, we used Spacer~\cite{DBLP:conf/cav/KomuravelliGCC13} to prove that the three implementation of the 2-bit counters behaves the same (i.e. each implementation outputs the stream true). 

%% file: conclu.tex

In this paper, we proposed a new compilation scheme to faithfully compile a mix of dataflow formalism and hierarchical state machines into modular Horn clauses. Our approach compile hierarchical state machines expressed as automata in the Lustre language into modular Horn clauses. The compilation technique preserves the structural and modal behavior of the system which makes the analysis of such models more tractable. The proposed approach is implemented in \lustrec  -- an open source Lustre compiler. Once the modular Horn clauses are generated, automated reasoning tools like Spacer~\cite{DBLP:conf/cav/KomuravelliGCC13} can be used to reason about properties. In the future, we plan to evaluate our compilation scheme on larger industrial case studies.



